\documentclass[12pt]{article}
\usepackage{epsfig}
\usepackage{feynmf}

\newcommand{\be}{\begin{eqnarray}}                   
\newcommand{\ee}{\end{eqnarray}}
\newcommand{\bc}{\begin{center}}
\newcommand{\ec}{\end{center}}

\title{\begin{flushright}
{\normalsize NUC-MINN-02/4-T\\
June 2002 \\
}
\end{flushright}
\vspace*{0.3in}
{\bf Dissipation at Two-Loop Level: \\ Undressing the Chiral Condensate}}
\author{{\bf \'Agnes M\'ocsy}$^{\star}$\\
{\it The Niels Bohr Institute,} \\
{\it Blegdamsvej 17, DK-2100 Copenhagen \O, Denmark}\\
{\it and}\\
  {\it School of Physics and Astronomy, University of Minnesota, }\\
   {\it Minneapolis, MN 55455, USA}}

\date{\today}

\begin{document}

\maketitle
\begin{fmffile}{work}
\begin{abstract}

A simple and consistent real time analysis of the long-wavelength chiral 
condensate fields in the background of hard thermal modes is presented in the 
framework of the linear sigma model. Effective evolution equations are 
derived for the inhomogeneous condensate fields coupled to a heat bath. The 
multiple effects of the thermal background on the disoriented chiral 
condensate are studied using linear response theory. We determine the 
temperature dependence of the equilibrium condensate, and examine the 
modification of the sigma and pion dispersion relations as these mesons 
traverse a hot medium. We calculate the widths by determining the dissipative 
coefficients at nonzero temperature at one- and two-loop order with resummed 
meson masses. Our results show that not only decay processes, but elastic 
scattering processes are significant at high temperatures, yielding to short 
relaxation times in the phase transition region. The relaxation times 
obtained are shorter than in previous estimates, making the observation of 
DCC signals questionable. Throughout this work Goldstone's Theorem is 
fulfilled when chiral symmetry is spontaneously broken. 

\end{abstract} 


\vspace*{0.2in}

\noindent
$^{\star}$ electronic address: mocsy@alf.nbi.dk\\

\section{Introduction \label{sect-intro}}

Non-equilibrium phenomena in many-body systems has received a great deal of 
attention during the last two decades. The theoretical interest keeps 
growing as new experiments are readily emerging in different research areas 
of physics. Such are the chiral phase transition and quark-gluon plasma in 
ultarrelativistic heavy ion collisions, ultrafast spectroscopy in 
semiconductors, Bose-Einstein condensates, strongly correlated Fermi 
systems in condensed matter physics, and electroweak baryogenesis and 
inflation in Early Universe cosmology. In the following we present 
a quantum field theoretical description of the dynamics of nuclear matter 
formed as a consequence of nuclear collisions at ultrarelativistic energies. 

The existence of a deconfined, chirally symmetric phase of QCD was predicted 
long ago \cite{laermann}. The interplay of results obtained from the three 
major research approaches, lattice simulations, theoretical models and
 experiments at RHIC-BNL, lead us to expect that 'finding' quark-gluon plasma 
is within reach. PHENIX results \cite{phenix} from the Year-1 run of RHIC 
suggest that the energy density achieved in Au+Au collisions already at 
$\sqrt{s}=130~$GeV is high enough to be favorable for the existence of free 
quarks and gluons. Early thermalisation of the partonic medium \cite{bjoraker} 
has been indicated by the elliptic flow analyses at STAR \cite{star}. However, 
most probably thermal equilibrium is not maintained as the matter rapidly
 expands and the temperature quickly drops through the QCD phase transition. 
Arguments that hint toward non-equilibrium, even explosive \cite{andy} 
dynamics in heavy ion collisions, such as large fluctuations of the average 
transverse momentum, and almost equal sideward and outward HBT radii, were 
reported in \cite{qm01}. The dynamical evolution of such an 
out-of-equilibrium  system is not yet understood.
When it comes to possible approaches, lattice simulations unfortunately do not 
prove to be the way to go: lattice can describe static situations in thermal 
equilibrium. Therefore, for learning about the dynamics of non-equilibrium 
systems, one has to reside to different theoretical approaches. Attempts are 
made using field theory \cite{boya96,cooper,rischke,magyar,cejk}, 
covariant kinetic theory \cite{molnar} and recently also non-equilibrium fluid 
dynamics \cite{muronga}. 

Until now no unambiguous observable that could serve as evidence for the 
phase transition has been identified. Based on the knowledge that a 
transition from ordered to disordered phase is accompanied by the formation of 
condensates, disoriented chiral condensates (DCC) have been suggested 
\cite{rw} as signature of the chiral phase transition. This means that the 
matter formed in a heavy ion collision can relax into a vacuum state that is 
oriented differently than the normal ground state. Relaxing of DCCs to the 
correct ground state then happens through the emission of low momentum pions 
with an anomalous distribution in isospin space. Detecting fluctuations in 
the ratio of produced neutral pions compared to the charged ones can serve as 
a signal for the deconfinement and chiral symmetry restoring phase 
transition. Such DCC signals have not been observed at CERN-SPS energies 
\cite{sps}. The STAR detector at RHIC searches for dynamical 
fluctuations on an event-by-event basis and may be therefore better suited 
for DCC searches. Other signals were proposed in context of DCCs: dileptons 
\cite{dileptons}, and recently the anomaly in the Omega and anti-Omega 
abundances observed at SPS \cite{omega}.

The ability to detect DCCs depends on their lifetime. In the original work 
\cite{rw} DCC formation was proposed with the assumption of a perfect quench. 
This means that after the critical temperature is reached the long-wavelength 
modes of the chiral condensate decouple from the thermal modes and evolve 
according to zero temperature equations of motion. A more realistic 
discussion, accounting for the presence of a thermalized background was first 
presented in \cite{birogreiner}. Due to possible energy exchange between 
different degrees of freedom dissipation occurs, which in turn reduces the 
lifetime of the condensate. This problem has been addressed previously in 
\cite{boya96,cooper,rischke,magyar,cejk,koch,rajagopal}. Most of the previous 
calculations though focus on the evolution of the order parameter only 
\cite{boya96,cejk}. Based on the observation that true pions have a finite 
width \cite{tytgat} the damping of pion condensates was considered 
\cite{rischke,magyar}, but only at one-loop level. The usual argument for 
neglecting two-loop contributions is that these are higher order in the 
coupling. However, there is an important reason for why two-loop 
contributions are not negligible: whereas decay processes can happen only 
under some kinematic conditions, scattering can always happen. Dissipation 
due to scattering has been evaluated for the scalar condensate in 
\cite{cejk}. A first attempt to look at the lifetime of DCCs based on 
scattering processes was presented in \cite{koch}. 

In this paper we present a quantum field theoretical analysis of homogeneous 
and inhomogeneous chiral condensate field configurations out of thermal 
equilibrium that are coupled to a thermal bath. The framework is the linear 
sigma model, which has proved to describe fairly well the hadronic phase of 
two-flavor QCD. We treat the long wavelength chiral fields classically and we 
account for the short wavelength fields perturbatively, but resumming certain 
diagrams. Such semi-classical description is acceptable since the occupation 
number of the low momentum modes is large.
The effect of the heat bath on the condensate is contained within the 
deviations of the thermal field fluctuations from their equilibrium values. 
We identify these deviations as time-delayed responses of the hard thermal 
modes to the presence of the condensate, and evaluate them 
using linear response theory, which provides the evolution of observables in 
real physical time. The response functions renormalize the equations of 
motion, modifying the particle properties, and give rise to dissipation. Due 
to possible interaction and thus energy exchange between the soft and the 
hard modes decay channels open up and particles can scatter. These processes 
are responsible for the dissipation of the condensate. To the best of our 
knowledge, a consistent incorporation of relaxation processes at one- and 
two-loop level has not been done before in this context. There is another 
important effect, that of the change in the velocity. We discuss this in 
\cite{future}.

It is important to determine the order of the chiral transition, as this 
influences the dynamical evolution of the system. Experimentally, 
large-acceptance detectors are now able to measure average as well as 
event-by-event observables, which in principle can distinguish between 
scenarios with a first order, a second order, or merely a crossover type of 
phase transition. QCD with two massless quarks possesses chiral symmetry, 
described by the $SU(2)_R\times SU(2)_L\simeq O(4)$ group. This continuous 
symmetry is spontaneously broken at low temperatures, resulting in the 
emergence of Goldstone bosons \cite{goldi}. Based on universality arguments 
\cite{pw} the transition between the chiral symmetric and symmetry broken 
phases is of second order. Nature, however, supplies a different situation, 
with non-zero quark masses. Small quark masses explicitly break chiral 
symmetry, which then alters the nature of the phase transition. The second 
order transition becomes a smooth cross-over, provided that the baryon 
density is zero. In case of a sufficiently large baryon chemical potential 
the transition is of first order, suggesting the existence of a tricritical 
point in the $(\mu_B,T)$ plane of the QCD phase diagram \cite{oveme}. Year-1 
RHIC results \cite{phenix} suggest that at $\sqrt{s}=130~$GeV the baryon 
density is approaching the low density limit. Expecting an even smaller 
baryon density at the maximum collision energy justifies us to work at finite 
temperature and zero baryon chemical potential. 

The paper is structured as follows: In Section \ref{sect-eom} we derive 
coarse-grained equations of motions for the long wavelength chiral condensate 
fields which are in in contact with a heat bath. In Section 
\ref{sect-response} the response of the thermal bath to the presence of the 
non-thermal condensate is evaluated. In Section \ref{sect-masses} we present 
the temperature-dependence of the equilibrium value of the condensate and 
that of the self-consistently calculated meson masses. It is important to 
emphasize that our analysis is performed with the fulfillment of Goldstone's 
Theorem for all temperatures below the critical one. Also, 
the tachyon problem appearing in the mean field treatments is eliminated. In 
Section \ref{sect-dissip} we analyze the dissipation of long-wavelength sigma 
and pion fields due to decay and scattering processes. In Section 
\ref{sect-time} the relaxation time of DCCs is determined. We summarize our 
results in the concluding Section \ref{sect-conclusions}.

\section{Deriving Equations of Motion \label{sect-eom}}

Within the linear sigma model framework the study of DCC formation and 
evolution is convenient, since the physical picture is rather transparent.
The theory is formulated in terms of the chiral field 
$\Phi =(\sigma,\vec{\pi})$. The scalar sigma field $\sigma$ describes the 
scalar quark condensate $\langle{\bar{q}}q\rangle$, which serves as order 
parameter. The pseudoscalar pion field $\vec{\pi} =(\pi^{1},\pi^{2},...,
\pi^{N-1})$ is directly related to the pseudoscalar condensate, 
$\langle{\bar{q}}{\vec{\tau}}\gamma_5q\rangle$. The dynamics of the chiral 
condensate is completely determined by the evolution equations in space and 
time for the long-wavelength chiral fields. In the following we derive 
effective equations for these low-momentum fields in a hard-momentum 
background.

The Lagrangian of the linear O(N) sigma model is
\be
L(\Phi) = \frac{1}{2}\partial_\mu \Phi\partial^\mu \Phi - U(\Phi)\,~,
\label{lag}
\ee
where the usual choice for the potential is 
\be
U(\Phi) = \frac{\lambda}{4}\left(\Phi^2 - v_0^2\right)^2 + H\sigma\, ~.
\ee
The explicit breaking of chiral symmetry is implemented through the H-term 
that tilts the potential in the sigma direction. $H = f_\pi m_\pi^2$ and 
$v_0^2 = f_\pi^2-m_\pi^2/\lambda$, where ${\it f}_{\pi}=93~$MeV is the pion 
decay constant, $\lambda$ is a positive dimensionless coupling constant, and 
$m_\pi = 138~$MeV is the zero temperature mass of the pion. The parameters of 
the Lagrangian are chosen such that for $H=0$ chiral symmetry is 
spontaneously broken in the vacuum. The potential then resembles the bottom 
of a wine bottle with the zero temperature minimum at $\bar\Phi=(f_\pi,0)$. 
The excitations in radial direction, the sigma mesons, have a mass 
$m_\sigma^2=2\lambda f_\pi^2$, and excitations along the azimuthal direction, 
the pions, are Goldstone bosons, with $m_\pi=0$.

The evolution of the fields in thermal equilibrium has been extensively 
studied \cite{baym,bochkarev} after the pioneering works by Linde 
\cite{linde} and Dolan and Jackiw \cite{dolan}. The usual procedure is to 
express the fields as 
\be
\sigma(t,\vec x) &=& v + \sigma_f(t,\vec x)\,~,\nonumber\\
\pi^i(t,\vec x) &=& \pi_f^i(t,\vec x)\,~, ~~i=1,...,N-1\,~.
\label{eq}
\ee
The thermal average of the chiral field, $\bar\Phi=(v,0)$, is the order 
parameter chosen to lie along the sigma direction, 
$\langle\sigma\rangle_{eq}=v$ and $\langle\vec\pi\rangle_{eq}=0$. The 
fluctuations about the order parameter average to zero, 
$\langle\sigma_f\rangle_{eq} = \langle\vec\pi_f\rangle_{eq} = 0$. Throughout 
this work we use the following notation:
\begin{itemize}
\item $\langle{\cal O}\rangle$ is the non-equilibrium ensemble average of an 
operator $\cal O~$;
\item $\langle{\cal O}\rangle_{eq}$ denotes the equilibrium, but interacting 
ensemble average;
\item $\langle{\cal O}\rangle_0$ is the free ensemble average. 
\end{itemize}

For non-equilibrium thermal conditions, the idea is that instead of writing the
fields as in (\ref{eq}), we separate them into their low and high frequency 
modes. This can be done, for example, by introducing a momentum cutoff 
$\Lambda_c$, as in \cite{bodecker}. Then, by integrating out the high 
frequency modes we obtain an effective theory for the low frequency, long 
wavelength modes. The fields can be separated as follows: 
\be 
\sigma(x) &=& \tilde{\sigma}(x) + \sigma_f(x)\nonumber\\
\pi^i(x) &=& \tilde{\pi}^i(x) + \pi_f^i(x)\,~, ~~i=1,...,N-1
\label{slow-fast}
\ee
\begin{itemize}
\item $\tilde{\sigma}$ and $\tilde{\pi}^i$ are slowly varying condensate 
fields, representing low frequency modes with momentum $\mid\vec{k}\mid<
\Lambda_c$. These soft modes are occupied by a large number of particles and 
may then be treated as classical fields;
\item $\sigma_f$ and $\pi_f^i$ are high frequency, fast modes with 
$\mid\vec{k}\mid>\Lambda_c$. These hard modes, representing quantum and 
thermal fluctuations, constitute a heat bath. 
\end{itemize}

A choice of $\Lambda_c=0$ would mean that only homogeneous condensates, with 
$\vec k=0$ momentum are studied. Here we discuss condensates that can also be  
inhomogeneous.

The problem to be solved, then, is to describe the evolution of long 
wavelength classical fields that are embedded into a thermal bath. In our 
approach, these soft modes follow classical equations of motion, whereas the 
effect of the hard thermal modes is taken into account in a perturbative 
manner. We are well aware of the problems of a perturbative treatment, since 
the sigma model is a strongly coupled effective theory: With the choice of 
for example $m_\sigma\simeq 600~$MeV for the vacuum mass of the sigma meson 
the coupling constant is $\lambda\simeq 20$. However, we improve the model by 
resumming a certain class of diagrams in the perturbation series. We believe 
that even if the self-consistent solutions are approximate only, they still 
yield to qualitatively reliable results. 

We average the Euler-Lagrange field equations over time and length scales that 
are short compared to the scales characterizing the change in the slow fields, 
but long relative to the scales of the quantum fluctuations. This is known as 
coarse-graining. The average of high frequency fluctuations is thus 
$\langle\sigma_f(x)\rangle = 0$ and $\langle\vec\pi_f(x)\rangle = 0$, while 
$\langle\tilde\sigma(x)\rangle = \tilde\sigma(x)$ and 
$\langle\tilde{\pi}^i(x)\rangle = \tilde{\pi}^i(x)$. It should be noted at 
this point that, compared to earlier works (for example \cite{cejk}), we 
allow for a nonzero ensemble average not only along the sigma direction, but 
also in the pion directions. In other words, we allow the formation of 
disoriented chiral condensates. Also, cross correlations between fluctuations 
of different fields are nonzero, $\langle\varphi_i\varphi_j\rangle\neq 0$, and 
moreover, we consider nonzero cubic fluctuations of the form 
$\langle\varphi_i\varphi_j\varphi_k\rangle\neq 0$ which arise at the two-loop 
level. Furthermore, we separate the non-equilibrium condensate fields:
\be
\tilde\sigma(x) &=& v + \sigma_s(x)\nonumber\\ 
\tilde\pi^i(x) &=& \pi_s^i(x)\,~,~~~i=1,...,N-1\,~.
\label{philow}
\ee
\begin{itemize}
\item $v$ is the equilibrium value of the chiral condensate chosen along the 
sigma direction;
\item $\sigma_s$ and $\pi_s^i$ are slow fluctuations about equilibrium.
\end{itemize}
The thermal equilibrium ensemble of the hard fluctuations is affected by the 
presence of the condensate. The full, non-equilibrium ensemble averages are 
basically two- and three-point functions of thermalized fields evaluated at 
the same space-time point. These can be written as the sum of the equilibrium 
ensemble average and a fluctuation about this:
\be
&&\langle\sigma_f^2\rangle = \langle\sigma_f^2\rangle_{eq} + 
\delta\langle\sigma_f^2\rangle \nonumber \\
&&\langle{\pi_f^i}^2\rangle = \langle{\pi_f^i}^2\rangle_{eq} + 
\delta\langle{\pi_f^i}^2\rangle \nonumber \\
&&\langle\sigma_f\pi_f^i\rangle = \delta\langle\sigma_f\pi_f^i\rangle 
\nonumber \\
&&\langle\pi_f^i\pi_f^j\rangle = \delta\langle\pi_f^i\pi_f^j\rangle 
\nonumber \\
&&\langle\sigma_f^3\rangle = \langle\sigma_f^3\rangle_{eq} + 
\delta\langle\sigma_f^3\rangle \nonumber\\
&&\langle\sigma_f{\pi_f^i}^2\rangle = \langle\sigma_f{\pi_f^i}^2\rangle_{eq} + 
\delta\langle\sigma_f{\pi_f^i}^2\rangle \nonumber\\
&&\langle\pi_f^i{\pi_f^j}^2\rangle = \delta\langle\pi_f^i{\pi_f^j}^2 \rangle 
\nonumber\\
&&\langle\sigma_f^2\pi_f^i\rangle = \delta\langle\sigma_f^2\pi_f^i\rangle\,~.
\label{fluctall}
\ee
The deviations in the fluctuations (of the general form 
$\delta\langle\varphi_i^n\varphi_j^m\rangle$) are the responses of the fast 
modes to the presence of slow $\sigma_s$ and $\vec\pi_s$ background fields. 
These responses are proportional to the slow fields raised to some positive 
power, and so they vanish, as they should, in the absence of the background. 
Notice the absence of $\langle\sigma_f\pi_f^i\rangle_{eq}$ and 
$\langle\pi_f^i\pi_f^j\rangle_{eq}$. Reason for this is that the correlation 
functions in equilibrium, in the absence of a background, are diagonal. On 
account of the cubic couplings, however, one finds nonzero 
$\langle\sigma_f^3\rangle_{eq}$ and $\langle\sigma_f{\pi_f^i}^2\rangle_{eq}$. 

In equilibrium, at fixed temperature, $\sigma_s=0$ and $\vec\pi_s=0$ and the 
equilibrium condensate satisfies the following equation: 
\be
\lambda v^3 + 3\lambda v\langle\sigma_f^2\rangle_{eq} + 
\lambda v\sum_{i=1}^{N-1}\langle{\pi_f^i}^2\rangle_{eq} + 
\lambda\langle\sigma_f^3\rangle_{eq} + 
\lambda\sum_{i=1}^{N-1}\langle\sigma_f{\pi_f^i}^2\rangle_{eq} 
- \lambda v_0^2v - H = 0 \,~.
\label{equilibm}
\ee
Since all components of the pion field are equivalent, we can write the sum 
as $\sum_{i=1}^{N-1}\langle{\pi_f^i}^2\rangle_{eq} = 
(N-1)\langle\pi_{f}^2\rangle_{eq}$, where now $\pi_{f}$ is one of the 
components. This is a convention that we adopt throughout the rest of the 
paper. 

The resulting final equations describing the evolution of the condensate 
fields are a set of coupled non-linear integro-differential equations. Let us 
assume only a small deviation from equilibrium, $|\sigma_s|, |\pi_s|\ll v$, 
and thus neglect terms that are higher order in $\sigma_s$ and $\pi_s$. The 
linearized field equations read 
\be
\partial^2\sigma_s + M_\sigma^2\sigma_s + 
\lambda v\left[3\delta\langle\sigma_f^2\rangle + 
(N-1)\delta\langle\pi_f^2\rangle\right] + 
\lambda\left[\delta\langle\sigma_f^3\rangle + 
(N-1)\delta\langle\sigma_f\pi_f^2\rangle\right] = 0 \,~,\nonumber
\ee
\be
M_\sigma^2 = \lambda\left(2v^2 - 
\frac{\langle\sigma_f^3\rangle_{eq}}{v} - 
(N-1)\frac{\langle\sigma_f\pi_f^2\rangle_{eq}}{v}\right)+ \frac{H}{v}\,~,
\label{sigmaeom2}
\ee
and
\be
\partial^2\pi_s + M_\pi^2\pi_s + 
2\lambda v\delta\langle\sigma_f\pi_f\rangle + 
\lambda\left[\delta\langle\pi_f^3\rangle + 
\delta\langle\sigma_f^2\pi_f\rangle\right] = 0\,~,\nonumber
\ee
\be
M_\pi^2 = \lambda\left(2\langle\pi_f^2\rangle_{eq} - 
2\langle\sigma_f^2\rangle_{eq} - \frac{\langle\sigma_f^3\rangle_{eq}}{v} - 
(N-1)\frac{\langle\sigma_f\pi_f^2\rangle_{eq}}{v}\right) + \frac{H}{v} \,~.
\label{pioneom2}
\ee
We denote the effective masses of the sigma meson and the pion by $M_\sigma$ 
and $M_\pi$, and their corresponding vacuum value by $m_\sigma$ and $m_\pi$, 
respectively. 

At one-loop order the effective equations become simpler:  
\be
\partial^2\sigma_s + M_\sigma^2\sigma_\sigma + 
\lambda v\left[3\delta\langle\sigma_f^2\rangle + 
(N-1)\delta\langle\pi_f^2\rangle\right] = 0 
\label{sigmaeomm}
\ee
\be
M_\sigma^2 = 2\lambda v^2 + \frac{H}{v}\,~,
\label{sigmamassh}
\ee
and
\be
\partial^2\pi_s + M_\pi^2\pi_s + 
2\lambda v\delta\langle\sigma_f\pi_f\rangle = 0 
\label{pioneomm}
\ee
\be
M_\pi^2 = \frac{H}{v} + \tilde m_\pi^2 = \frac{f_\pi}{v}m_\pi^2 + 
\tilde m_\pi^2 \,~,~~~~ \tilde m_\pi^2 = 2\lambda\left[
\langle\pi_f^2\rangle_{eq} -  \langle\sigma_f^2\rangle_{eq}\right]\,~,
\label{pionmassh}
\ee
and the equilibrium condensate satisfies 
\be
\lambda v^3 + \lambda\left[3\langle\sigma_f^2\rangle_{eq} + 
(N-1)\langle\pi_f^2\rangle_{eq}\right]v - \lambda v_0^2v - H = 0 \,~.
\label{equilib1loop}
\ee
%

\section{Non-Equilibrium Fluctuations \label{sect-response}}

Linear response theory is a very convenient tool when one is interested in 
monitoring the effect of an external field applied to a system initially 
in thermal equilibrium. When this effect is small a linear approximation is 
feasible. A response function expresses the difference between the 
expectation value of an operator before and after an external perturbation 
has been turned on. The effects of the soft condensate fields on the thermal 
medium are thus described by expressions of the form
\be
\delta\langle\sigma_f^n(x)\pi_f^m(x)\rangle = i\int_0^t dt'
\int d^3x'\langle\left[H_{perturbed}(x'),\sigma_f^n(x)\pi_f^m(x)\right]
\rangle_{eq}\,~.
\label{linresp}
\ee
Accordingly, we evaluate the commutators of different powers of the field 
operators at two separate space-time points in the fully interacting, but 
unperturbed ensemble. Initial conditions are set by the assumption that 
in heavy-ion collisions the system reaches a state of approximate local 
thermal equilibrium \cite{star}, then it cools while expanding out of 
equilibrium and reaches the critical temperature. We define the initial time 
$t=0$ by when the critical temperature is reached. This implies 
$\sigma_s(0,\vec{x})=0$ and $\vec\pi_s(0,\vec{x})=0$. The respective powers 
$n$ and $m$ are identified from the expressions in equations (\ref{sigmaeom2}) 
and (\ref{pioneom2}). Let us emphasize that these responses should be 
evaluated in the unperturbed, equilibrium ensemble, which does include all the 
interactions between the different modes. 

The possible couplings between low and high frequency modes are determined by 
evaluating the potential $U$, with the fields separated into their slow and 
fast components as above. The resulting Hamiltonian contains positive powers 
of the slow fields. For small departures from equilibrium it is enough to 
keep the dominant linear terms only. Relaxing the assumption of only slightly 
out of equilibrium requires the inclusion of higher powers of the non-thermal 
fields. This should provide no difficulties, but is beyond the aim of the 
present paper. The relevant couplings are
\be
&& H_{\sigma_s\sigma_f} = \lambda(3v^2-f_\pi^2)\sigma_s\sigma_f + 
3\lambda v\sigma_s\sigma_f^2 + \lambda\sigma_s\sigma_f^3\nonumber\\
&& H_{\pi_s\pi_f} = \lambda(v^2-f_\pi^2)\pi_s\pi_f + 
\lambda\pi_s\pi_f^3\nonumber\\
&& H_{\sigma_s\pi_f} = \lambda v\sigma_s\pi_f^2\nonumber\\
&& H_{\pi_s\sigma_f\pi_f} =  \lambda\pi_s\sigma_f^2\pi_f  + 
2 \lambda v\pi_s\sigma_f\pi_f\nonumber\\
&& H_{\sigma_s\sigma_f\pi_f} = \lambda\sigma_s\sigma_f\pi_f^2
\label{ham}
\ee
As before, $\pi_{s,f}$ refer to one of the $N-1$ components of the pion field. 

The response functions obtained by inserting (\ref{ham}) in (\ref{linresp}) are
\be 
\delta\langle\sigma_f^2(x)\rangle = 3i\lambda v\int_0^t dt'
\int d^3x'\sigma_s(x')\langle\left[\sigma_f^2(x'),\sigma_f^2(x)\right]
\rangle_{eq.} \,~,
\nonumber
\label{phi1}
\ee
\be 
\delta\langle\pi_f^2(x)\rangle = i \lambda v\int_0^t dt'\int d^3x' 
\sigma_s(x')\langle\left[\pi_f^2(x'),\pi_f^2(x)\right]\rangle_{eq.}\,~,
\nonumber
\label{phi2}
\ee
\be 
\delta\langle\sigma_f(x)\pi_f(x)\rangle = 2i\lambda v\int_0^t dt'
\int d^3x' \pi_s(x')\langle\left[\sigma_f(x')\pi_f(x'),\sigma_f(x)\pi_f(x)
\right]\rangle_{eq.} \,~.
\label{phi1phi2}
\ee
The cubic functions are determined analogously, therefore we skip presenting 
their evaluation. It is clear that expressions (\ref{phi1phi2}) vanish in the 
absence of the non-thermal background, as they should. 

The expectation values of the commutators are 
\be
\langle\left[\sigma_f^2(x'),\sigma_f^2(x)\right]\rangle_{eq.} &=& 
2 \left(D_\sigma^<(x,x')^2 - D_\sigma^>(x,x')^2\right) \,~,\nonumber\\
\langle\left[\pi_f^2(x'),\pi_f^2(x)\right]\rangle_{eq.} &=& 
2 \left(D_\pi^<(x,x')^2 - D_\pi^>(x,x')^2 \right) \,~,\nonumber\\
\langle\left[\sigma_f(x')\pi_f(x'),\sigma_f(x)\pi_f(x)\right]\rangle_{eq.} 
&=& D_\sigma^<(x,x')D_\pi^<(x,x') - D_\sigma^>(x,x')D_\pi^>(x,x') \,~,\nonumber
\label{expect}
\ee
where $x=(t,\vec x)$ and $x'=(t',\vec x')$ are four-vectors in coordinate
space. Keeping in mind that $t'<t$, where $t$ is the time elapsed after 
switching on the perturbation, and $t'$ is the time-variable that has its 
values in the $[0,t]$ interval, the following notation has been introduced:
\be
&& D_\sigma^>(x,x')\equiv \langle\sigma_f(x)\sigma_f(x')\rangle_{eq.}\,~, 
~~~ D_\pi^>(x,x')\equiv \langle\pi_f(x)\pi_f(x')\rangle_{eq.}\,~\nonumber\\
\mbox{and}\nonumber\\
&& D_\sigma^<(x,x')\equiv \langle\sigma_f(x')\sigma_f(x)\rangle_{eq.}\,~, 
~~~ D_\pi^<(x,x')\equiv \langle\pi_f(x')\pi_f(x)\rangle_{eq.}\,~.
\label{propag}
\ee
The functions $D^>$ and $D^<$ define the spectral function  
\be
\rho(k) &=& D^>(k)-D^<(k)\,~, 
\ee
which determines the real time propagator
\be
D(x,x') &=& \int\frac{d^4k}{(2\pi)^4}e^{-ik(x-x')}D(k)\nonumber\\
&=& \int\frac{d^4k}{(2\pi)^4}e^{-ik(x-x')}\left(\Theta(t-t')+f(k^0)\right)
\rho(k)\,~.
\label{relprop}
\ee
Here $k=(k^0,\vec{k})$ is the four-momentum, $f(k^0) = (e^{k^0\beta}-1)^{-1}$ 
is the Bose-Einstein distribution and $\beta = T^{-1}$. The response 
functions are then
\be
\delta\langle\sigma_f^2(x)\rangle &=& - i6\lambda v\int d^4x'\sigma_s(x')
\int\frac{d^4p}{(2\pi)^4}\int\frac{d^4q}{(2\pi)^4}e^{-i(p+q)(x-x')}\nonumber\\
&& \qquad \qquad \qquad\qquad\times 
\rho_\sigma(p)\rho_\sigma(q)(1+f(p^0)+f(q^0))\nonumber
\label{1^2}
\ee
\be
\delta\langle\pi_f^2(x)\rangle &=& - i2\lambda v\int d^4x'\sigma_s(x')
\int\frac{d^4p}{(2\pi)^4}\int\frac{d^4q}{(2\pi)^4} e^{-i(p+q)(x-x')}\nonumber\\
&& \qquad\qquad\qquad\qquad\times \rho_\pi(p)\rho_\pi(q)(1+f(p^0)+f(q^0))
\label{2^2}\nonumber
\ee
\be 
\delta\langle\sigma_f(x)\pi_f(x)\rangle &=& 
 - i2\lambda v\int d^4x'\pi_s(x')\int\frac{d^4p}{(2\pi)^4}
\int\frac{d^4q}{(2\pi)^4}e^{-i(p+q)(x-x')} \nonumber\\
&& \qquad \qquad\qquad \times \rho_\sigma(p)\rho_\pi(q)(1+f(p^0)+f(q^0))\,~.
\label{12}
\ee
%

\section{The Order Parameter and Meson Masses \label{sect-masses}}

To analyze the phase transition we look at the temperature dependence of the 
equilibrium condensate which is the order parameter of our model. The 
behavior of $v$ is determined by solving equation (\ref{equilibm}). It is
educational to look at the theory in the exact chiral limit, $H=0$, first. In 
this case equation (\ref{equilibm}) for the order parameter has two solutions:
\be
v =0\,~, ~~~~~ T>T_c\,~,\nonumber
\ee
and
\be
v^2 = f_\pi^2 - 3\langle\sigma_f^2\rangle_{eq} - 
(N-1)\langle\pi_f^2\rangle_{eq} - \frac{\langle\sigma_f^3\rangle_{eq}}{v} - 
(N-1)\frac{\langle\sigma_f\pi_f^2\rangle_{eq}}{v}\,~, ~~~T<T_c \,~.
\ee
The first solution shows, as we expect, that the chiral condensate does not 
exist above a critical temperature. The second solution represents the low 
temperature, symmetry broken phase. The two solutions are separated by a 
second order phase transition. 

The order parameter depends on the equilibrium field fluctuations, which to 
first approximation in either perturbative expansion in $\lambda$ 
\cite{kapusta} or $1/N$-expansion \cite{bochkarev}, are 
\be
\langle\varphi^2\rangle_{eq} = \int\frac{d^3p}{(2\pi)^3}
\frac{1}{2E}(1+2f(E))\,~,
\label{fluct}
\ee
where $\varphi=\sigma_f,\pi_f$ and $f$ is the Bose-Einstein distribution 
function of the mesons with energy $E=\sqrt{p^2+m^2}$. Cubic fluctuation 
terms that enter in the expression for the equilibrium condensate, 
$\langle\sigma_f^3\rangle_{eq}/{v}$ and 
$\langle\sigma_f\pi_f^2\rangle_{eq}/{v}$ are non-zero on account of the 
possible cubic couplings. However, already the leading order term of the 
expansion is higher order in the coupling than (\ref{fluct}), so we can drop 
these to first order.

Let us discuss some important observations regarding the fluctuations 
(\ref{fluct}). First, keeping only the leading order term means that the 
masses are the bare zero temperature masses. However, as indicated by 
equations (\ref{sigmamassh}) and (\ref{pionmassh}), the masses themselves are 
given in terms of the equilibrium condensate and as such, they are temperature 
dependent. Therefore, the temperature dependence of the order parameter should 
be determined by performing a self-consistent evaluation. Here we resum 
tadpole diagrams.

Second, the first term is the vacuum contribution, while the second term is 
due to finite temperature effects. The zero temperature part is divergent in 
the ultraviolet limit. This divergence can and should be removed using vacuum 
renormalization techniques. The finite temperature does not introduce any 
extra divergence since it is regularized by the distribution function. One 
can say that the very short distance behavior of the theory is not affected 
by finite temperature \cite{kapusta}. Therefore, $T=0$ renormalization is 
enough to obtain finite results. The usual approach then is to neglect the 
zero temperature contributions when focusing on the physics at finite 
temperatures. At this point, it is worth mentioning that using a 
self-consistent approximation makes the usual renormalization procedure 
difficult, as discussed in \cite{renorm}. The argument, according to which 
the renormalized divergent term can be neglected, is really correct only when 
the mass is the bare mass. A self-consistent calculation involves the 
temperature-dependent mass, leading to the temperature-dependence of the 
divergent term. Renormalization thus results in temperature-dependent 
renormalization constants, and these should not be ignored. However, such a 
treatment is beyond the scope of this paper, and in what follows, we are 
going to ignore the divergent term, while still being aware of this 
approximation.      

Third, the momentum integration has a lower limit, $\Lambda_c$, due to the 
restriction of the thermal population to hard momenta, $|\vec{p}|>\Lambda_c$. 
When $\Lambda_c=0$, the condensate contains only zero momentum modes, meaning 
that the classical field configurations are homogeneous. In reality non-zero 
but small momenta can be part of the condensate. Then we talk about 
inhomogeneous condensate.

\subsection{Goldstone Modes}

In the theory with spontaneously broken chiral symmetry the tree level mass 
of the pion is zero in the broken phase, $m_\pi=0$. Goldstone's Theorem 
\cite{goldi} requires that this remains zero at every order in perturbation 
theory. The first glimpse at equation (\ref{pioneom2}) (or (\ref{pionmassh})) 
 for the pion condensate shows a mass-term. At one-loop order  
\be
\tilde m_\pi^2 = 2\lambda\left[\langle\pi_f^2\rangle_{eq} - 
\langle\sigma_f^2\rangle_{eq}\right]\neq 0\,~,
\label{apparentmass}
\ee
which is zero only at i) zero temperature, where the thermal fluctuations 
themselves vanish, or ii) when the masses of the pion and the sigma are equal,
 which is expected at the critical temperature, or iii) in the high 
temperature limit, when the meson masses can be neglected. In the following, 
we present a simple and clear way to prove that this violation of Goldstone's 
Theorem is only apparent.

The pion mass (\ref{apparentmass}) includes one-loop tadpole contributions. 
There is another one-loop diagram that contributes to order $\lambda$, 'built' 
out of two 3-vertices, known as sunset diagram. This diagram is incorporated 
in the equation of motion through the response function 
$\delta\langle\pi_f\sigma_f\rangle$ given by (\ref{12}). The function is of 
the order of $\lambda v\sim\lambda^{1/2}$, and there is an overall factor of 
$2\lambda v\sim\lambda^{1/2}$ in front of it in the effective equation. 
Therefore, to get the contribution of order $\lambda$, the expectation value 
in (\ref{12}) can be evaluated at the lowest order. This means that we can 
replace the interacting ensemble average $\langle...\rangle_{eq.}$ by the 
free ensemble average $\langle...\rangle_0$. This in turn is equivalent to 
inserting free spectral functions,
\be
\rho_{free}(p) = 2\pi\epsilon(p^0)\delta(p^2-m^2)\,~,
\ee
into the expressions of (\ref{12}). After evaluating the frequency integrals 
we find
\be
\delta\langle\sigma_f\pi_f\rangle = i2\lambda M_\sigma^2\int 
d^4x'\pi_s(x')\int\frac{d^3p}{(2\pi)^3}\int\frac{d^3q}{(2\pi)^3} 
e^{i(\vec{p}+\vec{q})(\vec{x}-\vec{x}~')} F(\vec{p}, \vec{q}, t')\,~,
\ee
with
\be
F(\vec{p}, \vec{q}, t') &=& \frac{1}{4E_\sigma E_\pi}\left[(1+f_\sigma+f_\pi)
\left(e^{i(E_\sigma+E_\pi)(t-t')} -e^{-i(E_\sigma+E_\pi)(t-t')}\right)\right. 
\nonumber\\
&& \left. \qquad\qquad +(f_\pi-f_\sigma)\left(e^{i(E_\sigma-E_\pi)(t-t')}-
e^{-i(E_\sigma-E_\pi)(t-t')}\right)\right]\,.\nonumber\\
\label{F}\nonumber
\ee
Since, the deviation from equilibrium is assumed to be small, we Taylor expand 
 $\pi_s(x')$ about its equilibrium value. The first term of the expansion, 
linear in $\pi_s(x)$, gives the contribution to the mass. The total pion mass 
is then 
\be
M_\pi^2 = \tilde{m}_\pi^2 + a_1\,~,
\label{truemass}
\ee
where 
\be
a_1 &=& i2\lambda M_\sigma^2\int\frac{d^3p}{(2\pi)^3}\int_0^t dt'
F(\vec{p}, t')\nonumber\\
&=& i2\lambda M_\sigma^2\int\frac{d^3p}{(2\pi)^3}\frac{1}{4E_\sigma 
E_\pi}\left[\frac{2E_\sigma(1+2f_\pi)-2E_\pi(1+2f_\sigma)}{i(E_\pi^2-
E_\sigma^2)}\right] \nonumber\\
&=&2\lambda\int\frac{d^3p}{(2\pi)^3}\left[\frac{1}{2E_\sigma}(1+2f_\sigma) - 
\frac{1}{2E_\pi}(1+2f_\pi)\right] \,~,
\label{a1}
\ee
and $E_\sigma = \sqrt{\vec{p}~^2 + M_\sigma^2}$ and $E_\pi = |~\vec{p}~|$. 
Summing up (\ref{a1}) and (\ref{apparentmass}) in expression (\ref{truemass}), 
the true pion mass yields
\be
M_\pi = 0\,~,
\ee
thus proving that with proper inclusion of diagrams the pions stay Goldstone 
bosons at one-loop level at all temperatures.  Another simple and 
straightforward proof in frequency-momentum space is presented in 
\cite{future}.

\subsection{Numerical Results}

\begin{figure}[p] 
\begin {center}
\vspace*{-3cm}
\leavevmode
\hbox{%
\epsfysize=13cm
\epsffile{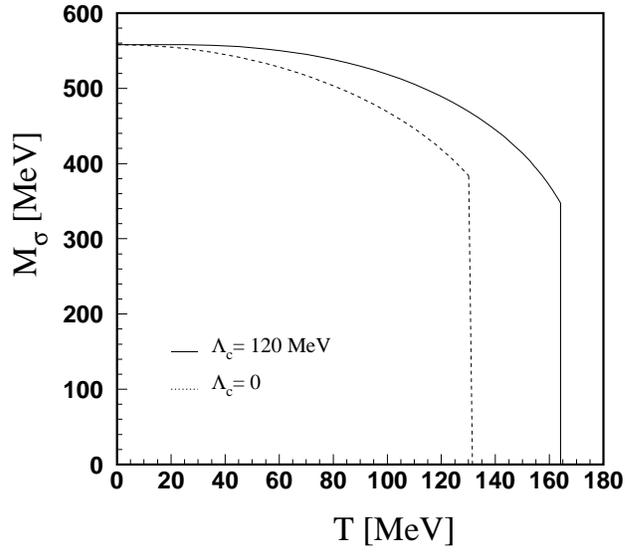}
}
\end{center}
\vspace*{-3.5cm}
\caption{Temperature dependence of the sigma mass for different momentum 
separation scales in the chiral limit.}
\label{lowlimmass.fig}
\end{figure}

In the exact chiral limit the self-consistent solution of the gap equation 
\be
v^2 = f_\pi^2 - 3\langle\sigma_f^2\rangle_{eq} - 
(N-1)\langle\pi_f^2\rangle_{eq} \,~,
\label{v}
\ee
with the sigma and pion field fluctuations 
\be
\langle\sigma_f^2\rangle = \frac{1}{2\pi^2}\int_{\Lambda_c}^\infty dp~
\frac{p^2}{E_\sigma}\frac{1}{e^{E_\sigma/T}-1}\,~,
\label{sigmafluct}
\ee
and 
\be
\langle\pi_f^2\rangle = \frac{1}{2\pi^2}\int_{\Lambda_c}^\infty dp~
\frac{p^2}{E_\pi}\frac{1}{e^{E_\pi/T}-1} \rightarrow \frac{T^2}{12},~~~~
\mbox{for}~ M_\pi\rightarrow 0\,~, \Lambda_c\rightarrow 0\,~,
\label{pionfluct}
\ee
evaluated with $E_\sigma=\sqrt{p^2+M_\sigma^2}$, where
\be
M_\sigma^2 = 2\lambda v^2\,~,
\label{sigmam}
\ee 
is presented in figure \ref{lowlimmass.fig} for different values of the 
momentum separation scale $\Lambda_c$. In the numerical analysis $N=4$ and the 
coupling constant was chosen to be $\lambda=18$, corresponding to a vacuum 
sigma mass of about $m_\sigma^2 = 2\lambda f_\pi^2 = (558~$MeV$)^2$.
\begin{figure}[p] 
\begin {center}
\vspace*{-3cm}
\leavevmode
\hbox{%
\epsfysize=13cm
\epsffile{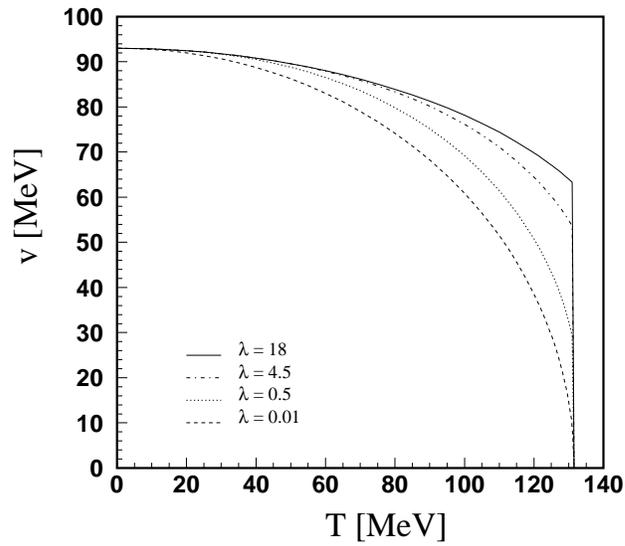}
}
\end{center}
\vspace*{-3.5cm}
\caption{Temperature dependence of the equilibrium condensate for different 
coupling constants in the chiral limit.}
\label{lambda.fig}
\end{figure} 
The phase transition temperature, defined by the vanishing of the condensate, 
is about $T_c\simeq 130~$MeV for homogeneous classical field configurations, 
$\Lambda_c = 0$. With the increase of the scale $\Lambda_c$ the phase-space 
available for hard modes is decreased, requiring higher temperatures for the 
fluctuations to completely dissolve the condensate. For $\Lambda_c = 120~$MeV 
the temperature above which fluctuations are much too large to allow the 
formation of any condensate is about $T_c\simeq 165~$MeV, which is in good 
agreement with lattice data \cite{karsch}. Figure \ref{lowlimmass.fig} shows 
that the meson mass is positive-definite at all temperatures, eliminating the 
tachyon problem present in the mean field approximations \cite{kapusta}.

\begin{figure}[h] 
\begin {center}
\vspace*{-2cm}
\leavevmode
\hbox{%
\epsfysize=13cm
\epsffile{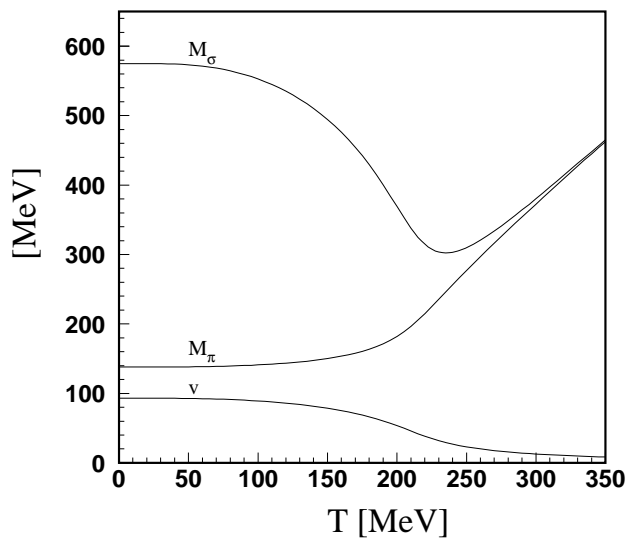}
}
\end{center}
\vspace*{-3.5cm}
\caption{Temperature dependence of the resummed meson masses, and of the 
equilibrium condensate in the $O(4)$ model with explicitly broken symmetry.}
\label{step.fig}
\end{figure} 

The theory with exact chiral symmetry is known to have a second order phase 
transition based on universality arguments \cite{pw}. This has been confirmed 
within the mean field approximation \cite{bochkarev}. Figure 
\ref{lowlimmass.fig} however, shows a discontinuous behavior of the order 
parameter at $T_c$. Such a jump is characteristic to first-order phase 
transitions. Incorporating the effect of thermal fluctuations in a 
self-consistent way renders the transition first order. Such a behavior has 
been discussed many years ago also by Baym and Grinstein \cite{baym}. In 
figure \ref{lambda.fig} we present how the discontinuity decreases with 
decreasing coupling constant. For a weak enough coupling the continuous 
second-order transition is recovered. This can be understood in the following 
way: in order to assure the finiteness of the $O(N)$-theory in the large-$N$ 
limit the coupling constant should be written as $\lambda/N$ \cite{baym} in 
the self-consistent gap equation (\ref{sigmam})-(\ref{v}):
\be
M_\sigma^2 = m_\sigma^2 - 
2\frac{\lambda}{N}\left[3\langle\sigma_f^2\rangle_{eq} + 
(N-1)\langle\pi_f^2\rangle_{eq}\right] \,~.
\label{sm}
\ee
For $N\rightarrow\infty$ the contribution from the sigma field fluctuation 
disappears. The condensate is then
\be
v^2 = f_\pi^2-\frac{N-1}{N}\langle\pi_f^2\rangle_{eq} \simeq 
f_\pi^2 - \frac{T^2}{12}
\ee
clearly showing a continuous decrease of the order parameter with increasing 
temperature. In the $O(4)$ model decreasing $\lambda$ by hand is equivalent 
to going to the large $N$ limit in the $O(N)$ model. For our model parameters 
we have determined a large $N_{critical}\simeq 1800$ at which the transition 
is second order. For this $\lambda=18$ was held fixed. This result 
is equivalent to having $\lambda_{critical}\simeq 0.01$ and $N=4$.

Numerically determined self-consistent solutions for the condensate 
(\ref{equilib1loop}) and the meson masses (\ref{sigmamassh}) and 
(\ref{pionmassh}) in the more realistic theory with explicitly broken 
symmetry are displayed in figure \ref{step.fig}, and show a qualitatively 
different behavior. There is no phase transition in the textbook sense. The 
equilibrium condensate monotonically decreases with increasing temperature. A 
crossover region can be defined where the 
sigma and pion masses start to approach degeneracy. The minimum of the sigma 
mass is at $T\simeq 235~$MeV. Different values for $\Lambda_c$ do not 
introduce significant effect in evaluating the meson masses.

\section{Dissipation of the Chiral Condensate \label{sect-dissip}}

Dissipation of the condensate occurs because energy can be transferred 
between the condensate and the heat bath through the interactions of soft and 
hard degrees of freedom. Formally, in our model, the damping of different 
modes is determined from the response functions. Most of previous studies on 
this topic have been done at one-loop level. The two-loop level scattering 
processes have been ignored, based on the argument that these are higher 
order in the coupling constant than are the decay and absorbtion processes. 
However, as we show in the following, two-loop processes can dominate due to 
the available large phase space.   

The effective equations of motion for long-wavelength meson fields 
(\ref{sigmaeomm}) and (\ref{pioneomm}) have the general form
\be
\partial^2\phi_s(x) + M^2 \phi_s(x) + F(x) = 0 \,~,
\label{geneom}
\ee
where $\phi_s=\sigma_s$ or $\pi_s$ and  
\be
F(x) = \int d^4x'\phi_s(x')\Pi(x-x')\,~.
\ee
In frequency-momentum space this reads as
\be
-k^2 + M^2 + \Pi(k) = 0\,~.
\label{keom}
\ee
The function $\Pi(k)$ is given by
\be
\Pi(k) &=& - ig^2\int d^4x'\int\frac{d^4p}{(2\pi)^4}\int\frac{d^4q}{(2\pi)^4}
e^{i(k-p-q)(x-x')}\nonumber\\
&&\qquad\qquad\qquad\qquad\times\rho_1(p)\rho_2(q)(1+f(p^0)+f(q^0))\,~,
\label{integral}
\ee
and is identified as the self-energy. Here $g$ is the corresponding coupling, 
and the indices $1$ and $2$ refer to 
either of the hard modes, $\sigma_f$, and $\pi_f$, respectively, and 
$k=(k^0,\vec{k})$ is the four-momentum of the soft sigma meson or pion. The 
frequency has a real and an imaginary part, $k^0=\omega-i\Gamma$, provided 
$\vec{k}$ real. The real part of the self-energy participates in the 
dispersion relation 
\be
\omega^2 = \vec{k}~^2 + M^2 + \mbox{Re}~\Pi(\omega,\vec{k})\,~,
\ee
and with the usual assumption of weak damping, $\Gamma\ll\omega$, the 
imaginary part of the self-energy completely determines the damping of 
excitations: 
\be
\Gamma = -\frac{\mbox{Im}~\Pi(\omega,\vec{k})}{2\omega}\,~.
\label{Gamma/2}
\ee
$\Gamma$ is the rate at which out-of-equilibrium meson modes with energy 
$\omega$ and momentum $\vec{k}$ approach equilibrium. This rate is determined 
by physical processes which can be identified from the imaginary part of the 
self-energy \cite{weldon}. One should be aware that (\ref{Gamma/2}) 
describes the rate of decay of the amplitude of the wave, $\exp(-\Gamma t)$. 
The loss rate for the number density is $\exp(-2 \Gamma t)$.

\subsection{Dissipation at One-Loop Order\label{sect-dissip1}}

There are several diagrams contributing to the self-energy at one-loop order.
Tadpoles are real and they only modify the mass, as we discussed before. 
Dissipative effects come from non-local diagrams. At order $\lambda$ these are 
determined from the response functions through (\ref{integral}) with the 
insertion of the free spectral functions,
\be
\Pi(k)\!\!\! &=& \!\!\!\!g^2\int\frac{d^3p}{(2\pi)^3} 
\frac{1}{4E_1E_2}\left[(1+f_1+f_2)
\left(\frac{1}{\omega-E_1-E_2+i\epsilon}-\frac{1}{\omega+E_1+E_2 + i\epsilon}
\right)\right.\nonumber\\
&& \qquad\qquad\qquad \left. 
+ (f_2-f_1)\left(\frac{1}{\omega-E_1+E_2+i\epsilon} - 
\frac{1}{\omega+E_1-E_2+i\epsilon}\right)\right]\,~, \nonumber
\ee
with $E_1=\sqrt{m_1^2+(\vec{p}+\vec{k})^2}$ and $E_1=\sqrt{m_2^2+\vec{p}~^2}$. 
The indices $1$ and $2$ refer again to either of the fast sigma and pion. It 
is important to observe that the above expression coincides with the 
self-energy calculated directly from the non-local one-loop diagram 
\cite{weldon}. For positive energies of the mesons $\omega \geq 0$ the 
contributions to the imaginary part of the self-energy are 
\be
&&\!\!\!\!\!\!\!\!\!\!\!\mbox{Im}~\Pi(\omega,\vec{k}) = 
-\pi g^2\int\frac{d^3p}{(2\pi)^3}\frac{1}{4E_1E_2}
\left[(1+f_1+f_2)\delta(\omega-E_1-E_2) \right.\nonumber\\
&&\left. \qquad\qquad \qquad\qquad\qquad\qquad\qquad\qquad + 
(f_2-f_1)\delta(\omega-E_1+E_2) \right]\,\, .
\label{relevantgen}
\ee
Because the heat bath singles out a preferred reference frame we keep 
$\omega$ and $\vec k$ independent. The dynamics of the decay is determined by 
the on-shell processes that are allowed.  

\begin{figure}[htbp]
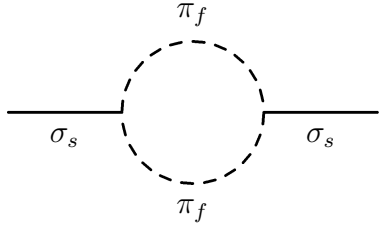

 \bc
 ~\parbox{6cm}{\begin{fmfchar*}(140,125)
   \fmfleft{i}
   \fmfright{o}
   \fmf{plain,label=$\sigma_s$}{i,v1}
   \fmf{plain,label=$\sigma_s$}{v2,o}
   \fmf{dashes,left,tension=.4,label=$\pi_f$}{v1,v2,v1}  
 \end{fmfchar*}}

 \parbox{16cm}{\caption{\small One-loop self-energy contribution to the soft 
$\sigma_s$ with coupling to the hard $\pi_f$ through 
$\lambda v\sigma_s\pi_f^2$.
 \label{si.fig}}}
 \ec
 \end{figure}

\paragraph*{Sigma Meson Decay.}
Contribution to the imaginary part of the sigma meson self-energy 
comes from the diagram presented in figure \ref{si.fig}. 
 \noindent
Kinematically, the decay of a soft sigma meson into hard thermal pions, 
$\sigma_s\rightarrow\pi_f\pi_f$, and the inverse, recombination 
process is allowed, provided that $\omega^2-k^2\geq 4m_\pi^2$.
\be
\mbox{Im}~\Pi(\omega,\vec{k}) = -\frac{g^2}{16\pi}\left[\sqrt{1-\frac{4m_\pi^2}
{\omega^2-k^2}} + 2\frac{T}{k}\log\left(\frac{1-e^{-\frac{\omega_{+}}{T}}}
{1-e^{-\frac{\omega_{-}}{T}}}\right)\right]\,\, ,
\ee
where
\be
\omega_{\pm} = \frac{\omega}{2}\pm \frac{k}{2}
\sqrt{1-\frac{4m_\pi^2}{\omega^2-k^2}}\,\, .\nonumber
\ee
The decay and formation processes are obtained by setting $\omega^2-k^2 = 
m_\sigma^2$ 
\be
\mbox{Im}~\Pi(m_\sigma) =  -\frac{\lambda^2 v^2}{16\pi}
\left[\sqrt{1-\frac{4m_\pi^2}
{m_\sigma^2}} + 2\frac{T}{k}\log\left(\frac{1-\exp{(-\frac{\omega}{2T}-
\frac{k}{2T}\sqrt{1-\frac{4m_\pi^2}{m_\sigma^2}})}}{1-\exp{(-\frac{\omega}{2T}
+\frac{k}{2T}\sqrt{1-\frac{4_\pi^2}{m_\sigma^2}})}}\right)\right]\,\, .
\ee
The rate at which soft, $k\ll T$, sigmas of energy $\omega$ disappear from 
the condensate due to their decay into thermal pions is
\be
\Gamma_{\sigma\pi\pi}(\omega) =  \frac{(N-1)}{16\pi}\lambda
\frac{m_\sigma^2-m_\pi^2}{\omega}\sqrt{1-\frac{4m_\pi^2}{m_\sigma^2}}
\coth{\left(\frac{\omega}{4T}\right)}\,\, .
\label{sipipi}
\ee

\begin{figure}[htbp]
\begin {center}
\vspace*{-4cm}
\leavevmode
\hbox{%
\epsfysize=14cm
\epsffile{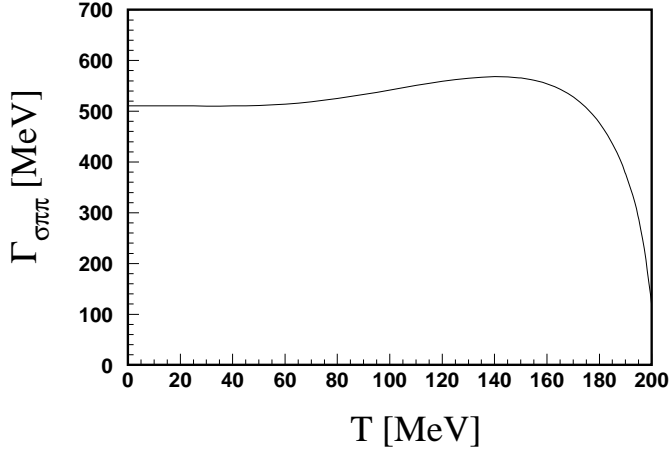} 
}
\vspace*{-4cm}
\caption{Temperature dependence of the sigma damping rate at one-loop order 
in the sigma rest frame. Calculations were done with resummed meson masses.
\label{sidecay.fig}}
\end{center}
\end{figure}

The temperature dependence of the sigma damping rate (\ref{sipipi}) in the 
rest frame of the sigma, $\omega=m_\sigma$, is shown in figure 
\ref{sidecay.fig}. The calculations were done with the resummed meson masses, 
$M_\sigma$ and $M_\pi$. Figure \ref{sidecay.fig} shows that even at zero 
temperature there is a finite damping, so the sigma meson can decay into two 
pions in the vacuum. At $T=0$, for our model parameters the damping rate is 
about $\Gamma_{\sigma\pi\pi}=510~$MeV, which is of the order of the mass of 
the sigma. The width of the sigma resonance is very broad, in other words the 
sigma meson is overdamped. Figure \ref{sidecay.fig} shows that the damping is 
increasing with $T$, and is followed by a quite drastic decrease starting from 
about $T=150~$MeV. This temperature corresponds to the temperature where the 
sigma mass begins to drop significantly (see figure \ref{step.fig}). It is 
then natural to expect a decrease of the sigma decay rate into pions that 
have masses approaching that of the sigma. Above $T\simeq 200~$MeV the 
threshold condition for the decay of an on-shell sigma meson, 
$M_\sigma\geq 2M_\pi$, is not fulfilled anymore. Therefore, there will be no 
contribution from decay to the sigma width in the kinematically suppressed 
region. 
%

\begin{figure}[h]
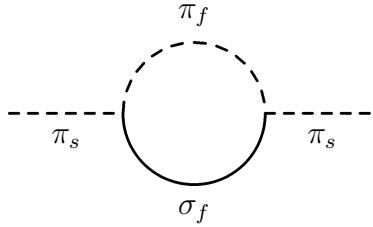

 \bc
 \begin{fmfchar*}(140,125)
   \fmfleft{i}
   \fmfright{o}
   \fmf{dashes,label=$\pi_s$}{i,v1}
   \fmf{dashes,label=$\pi_s$}{v2,o}
   \fmf{dashes,left,tension=.4,label=$\pi_f$}{v1,v2}  
   \fmf{plain,left,tension=.4,label=$\sigma_f$}{v2,v1}
 \end{fmfchar*}
 
 \parbox{16cm}{\caption{\small One-loop self-energy contribution to $\pi_s$ 
 with coupling to $\sigma_s$ and $\pi_s$ through $2\lambda v\pi_s
 \sigma_f\pi_f$.
 \label{pi.fig}}}
 \ec
 \end{figure}

\begin{figure}[h]
\begin {center}
\vspace*{-3cm}
\leavevmode
\hbox{%
\epsfysize=15cm
\epsffile{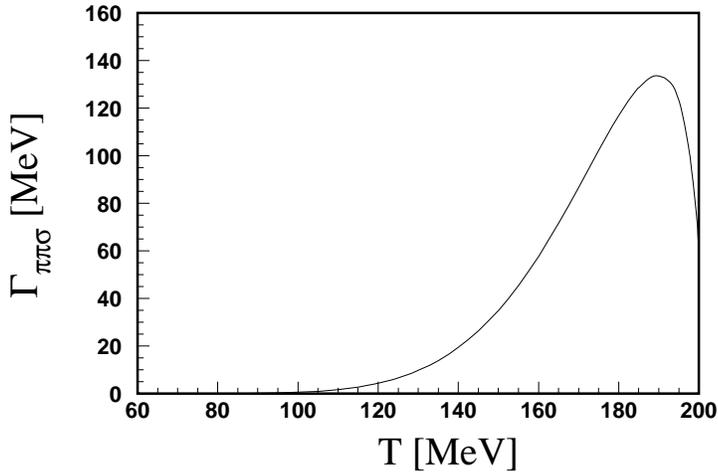} 
}
\vspace*{-4.3cm}
\caption{Temperature dependence of the pion damping rate at one-loop order in 
the pion rest frame. Calculations were done with resummed meson masses. 
\label{pidecay.fig}}
\end{center}
\end{figure}

\paragraph*{Pion Damping.}
At one-loop order there is only one diagram contributing to the pion 
self-energy. This diagram is shown in figure \ref{pi.fig}. At this order, 
dissipation of the pion condensate can occur provided the energy and momentum 
of the soft pion satisfies the kinematic condition 
$\omega^2-k^2\leq (m_\sigma-m_\pi)^2$. Then the transformation of a pion 
into a sigma when propagating through a thermal medium can happen.
 \noindent
Basically, 
a soft pion from the condensate annihilates with a hard thermal pion 
producing a hard thermal sigma meson, $\pi_s\pi_f\rightarrow\sigma_f$. The 
inverse process is the decay of a hard thermal sigma meson into a soft and a 
hard pion, $\sigma_f\rightarrow\pi_s\pi_f$. The net rate of dissipation is
\be
\Gamma_{\pi\pi\sigma}(\omega,\vec{k}) =  \frac{1}{8\pi}\lambda
\frac{T(m_\sigma^2-m_\pi^2)}{k\omega}\left[\log{\left(\frac{e^{\frac{\omega_+}
{T}}-1}{e^{\frac{\omega_-}{T}}-1}\right)} + \log{\left(\frac{e^{
\frac{\omega_+ +\omega}{T}}-1}{e^{\frac{\omega_- +\omega}{T}}-1}\right)}\right]
\label{pipisi}
\ee
where
\be
\omega_\pm &=& \sqrt{p_\pm^2+m_\pi^2}\, ,\nonumber\\
p_\pm &=& \pm\frac{k}{2}\frac{m_\sigma^2-2m_\pi^2}{m_\pi^2} + \frac{\omega}{2}
\frac{m_\sigma^2}{m_\pi^2}\sqrt{1-\frac{4m_\pi^2}{m_\sigma^2}} \,~.
\nonumber
\ee
In the rest-frame of the massive pion the expression for the damping is 
simplified to
\be
\Gamma_{\pi\pi\sigma}(m_\pi) =  \frac{\lambda}{16\pi}\frac{m_\sigma^2(m_\sigma^2 - 
m_\pi^2)}{m_\pi^3}\sqrt{1-\frac{4m_\pi^2}{m_\sigma^2}}
\frac{1-e^{-\frac{m_\pi}{T}}}{(e^\frac{m_\sigma^2-2m_\pi^2}{2m_\pi T}-1)
(1-e^{-\frac{m_\sigma^2}{2m_\pi T}})}\,~.
\label{pipis}
\ee
Figure \ref{pidecay.fig} shows the temperature dependence of the damping rate 
of massive pions at one-loop order (\ref{pipis}) calculated in the pion's rest 
frame using the resummed meson masses, $M_\sigma$ and $M_\pi$. Note that at 
zero temperature the dissipation is zero. This makes sense because the 
transformation of pions into sigmas is due to their annihilation with a hard 
thermal pion in the medium, and so this is exclusively a finite temperature 
process. 
At low temperatures the phase space available for this pion 
transformation process is suppressed by the large sigma mass. As the 
temperature increases and the sigma mass is dropping the width of the pions 
is increasing. Figure \ref{pidecay.fig} shows that the damping can get quite 
strong. At $T\simeq 170~$MeV, for example, when the pion mass is about 
$m_\pi=158~$MeV the damping is $\Gamma_{\pi\pi\sigma}\simeq 87.0~$MeV. Thus, 
at this temperature the width of the pion is about $55\%$ of its energy. 
This result makes us question whether the pion is a quasiparticle in this 
temperature region and needs further investigations. At temperatures around 
$200~$MeV the kinematic condition for an on-shell pion, $M_\sigma\geq 2M_\pi$, 
is not satisfied, prohibiting the transformation of a pion into a sigma while 
passing through a hot medium.  
\begin{figure}[h]
\begin {center}
\vspace*{-3cm}
\leavevmode
\hbox{%
\epsfysize=19cm
\epsffile{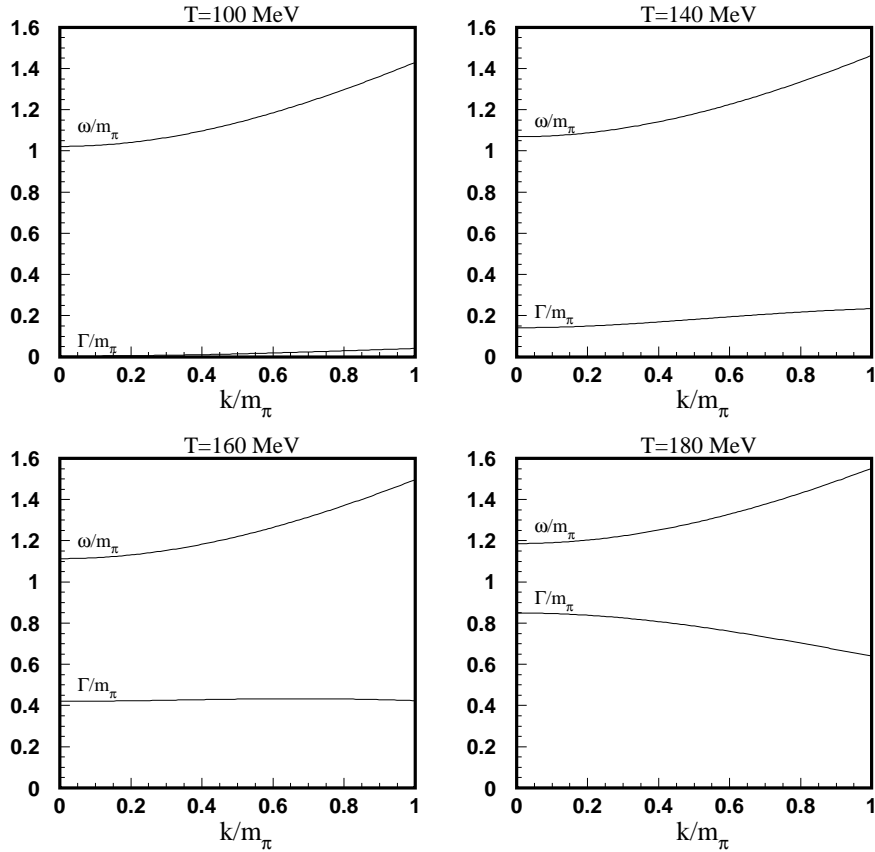} 
}
\vspace*{-4.5cm}
\caption{Momentum dependence of the pion energy and width at different 
temperatures calculated with resummed meson masses and with dispersion 
relation $\omega^2=k^2+M_\pi^2$. All quantities are normalized to the vacuum 
pion mass. 
\label{hdisipk.fig}}
\end{center}
\end{figure}

The damping in terms of the pion momenta is presented in figure 
\ref{hdisipk.fig} for different temperatures. Results using the dispersion 
relation $\omega^2 = k^2+M_\pi^2$ are displayed. Noticeable damping occurs 
above $T=100~$MeV and increases with $T$. At about $T=160~$MeV all modes are 
equally damped. In other words, the width of the pion is independent of 
its momentum. This width is increasing with $T$ and it can be as great as 
$30\%$ of the energy. At even higher temperatures the damping of the zero 
momentum modes is the strongest. 

\subsection{Dissipation at Two-Loop Order \label{sect-dissip2}}

Instead of the tedious evaluation of two-loop linear response functions, we 
present a direct determination of damping rates from the physical processes 
responsible. Two-loop contributions correspond to two-particle scatterings 
with amplitudes evaluated at tree level. Similar to the one-loop calculations 
we evaluate the imaginary parts of the two-loop self-energies and inserting 
these into (\ref{Gamma/2}) results in the damping rate due to scattering.  

The general form of the self-energy of a particle of mass $m_a$, propagating 
with four-momentum $k=(\omega,\vec{k})$ through a medium in thermal 
equilibrium, is given by \cite{shuryak} 
\be
\Pi_{ab}(k) = \int\frac{d^3p}{(2\pi)^32E} f(E){\cal M}(s)\,~.
\label{shuryak0}
\ee
Here ${\cal M}$ is the transition amplitude for the scattering process 
$ab\rightarrow ab$. The thermodynamical weight $f(E)$ is the Bose 
distribution of thermal mesons of mass $m_b$ and four-momentum 
$p=(E,\vec{p})$. In terms of the forward scattering amplitude 
${\cal M}(s)= -8\pi\sqrt{s}f_{cm}(s)$, where $s=(p+k)^2$, and the imaginary 
part follows
\be
\mbox{Im}~\Pi_{ab}(k) &=& -\int\frac{d^3p}{(2\pi)^3}f(E)\sqrt{s}
\frac{q_{cm}}{E}\sigma_{total}(s)\,~.
\ee
To obtain this equality we applied the standard form of the optical theorem 
that relates the imaginary part of the forward scattering amplitude and the 
total cross-section \cite{peskin}:
\be
\mbox{Im}~f_{cm}(s) = \frac{q_{cm}}{4\pi}\sigma_{total}(s)\,~.
\ee
Here we consider scatterings involving massive particles only. Then 
\be
\mbox{Im}~\Pi_{ab}(\omega) = -\frac{1}{8\pi^2}\int_{-1}^{1}d\cos{\theta}
\int_{\Lambda_c}^{\infty}dp\frac{p^2}{E}f(E)\sigma_{ab}(E)
\sqrt{(s-m_b^2+m_a^2)^2-4sm_a^2} \,~,\nonumber
\ee
with $s=m_a^2+m_b^2+2E\omega-2pk\cos{\theta}$, where $\theta$ is the angle 
between $\vec{k}$ and $\vec{p}$. The dispersion relation of the hard thermal 
modes is $E=\sqrt{p^2+m_b^2}$ and of the mesons inside the condensate 
$\omega=\sqrt{k^2+m_a^2}$. It is convenient to work in the rest frame of 
meson $a$, where
\be
\mbox{Im}~\Pi_{ab}(\omega=m_a,\vec{k}=0) = -\frac{m_a}{2\pi^2}
\int_{m_b}^{\infty}dE(E^2-m_b^2)f(E)\sigma_{ab}(E)\,~.
\ee
The cross section for a scattering $ab\rightarrow ab$ is given by
\be
\sigma_{ab} = \frac{1}{S!}\int\left(\frac{d\sigma_{ab}}{d\Omega}\right)_{cm}
d\Omega
\label{cross}
\ee
where
\be
\left(\frac{d\sigma_{ab}}{d\Omega}\right)_{cm} = \frac{1}{64\pi^2s}
|{\cal M}|^2\,~.
\label{dcross}
\ee
The symmetry factor $1/S!$ is due to the number $S$ of identical particles in 
the final state. It is clear that knowing the amplitude ${\cal M}$ of a 
process readily results in the dissipation rate due to that process.

\paragraph*{Sigma Scattering.}
There are two mechanisms that contribute to the removal or addition of a 
sigma meson to the condensate: elastic scattering of a hard thermal sigma or 
of a hard thermal pion off a sigma from the condensate. To first order in the 
coupling $\lambda$ there are four diagrams contributing to the process in 
which a thermal sigma meson knocks out a low momentum sigma from the 
condensate. The transition amplitude, the sum of contributions from different 
diagrams is: 
\be 
{\cal M} &=& -6\lambda\left[1+3(m_\sigma^2-m_\pi^2)\left(
\frac{1}{s-m_\sigma^2} + \frac{1}{t-m_\sigma^2} +
\frac{1}{u-m_\sigma^2}\right)\right]\,~,
\ee
This reflects the symmetry in the $s$, $t$, and $u$ channels, and 
$s+t+u= 4m_\sigma^2$. The total cross-section is 
\be
\sigma_{\sigma\sigma}(s) &=& \frac{9\lambda^2}{8\pi s}\left[\left(
\frac{s+2m_\sigma^2-3m_\pi^2}{s-m_\sigma^2}\right)^2 + 
\frac{18(m_\sigma^2-m_\pi^2)^2}{m_\sigma^2(s-3m_\sigma^2)} \right. \nonumber\\
&-& \!\!\!\! \left.\frac{12(m_\sigma^2-m_\pi^2)(s^2-3sm_\sigma^2-
m_\sigma^4+3m_\sigma^2m_\pi^2)}{(s-4m_\sigma^2)(s-2m_\sigma^2)
(s-m_\sigma^2)}\ln\left(\frac{s-3m_\sigma^2}{m_\sigma^2}\right)\right]
\ee
In the low energy limit expand this about $s=4m_\sigma^2$. To leading order  
\be
\sigma_{\sigma\sigma} = \frac{9}{32\pi}\lambda^2\frac{(4m_\sigma^2-
5m_\pi^2)^2}{m_\sigma^6} \,~,
\ee
and in the rest-frame of the sigma this gives rise to
\be
\mbox{Im}~\Pi_{\sigma\sigma} = -\frac{9}{32\pi^3}\lambda^2T^2e^{-m_\sigma/T}
\frac{(m_\sigma+T)(4m_\sigma^2-5m_\pi^2)^2}{m_\sigma^5}\,~.
\ee
In the high energy limit only the four-point vertex contributes to the 
amplitude, resulting in
\be
\sigma_{\sigma\sigma} = \frac{9\lambda^2}{8\pi s}\,~.
\ee
This gives rise to
\be
\mbox{Im}~\Pi_{\sigma\sigma} = -\frac{3}{64\pi}\lambda^2T^2\,~.
\ee
With these two limits we construct an interpolating formula that describes 
the whole energy range. The contribution to the rate of decay of the 
amplitude is then 
\be
\Gamma_{\sigma\sigma} \simeq \frac{9\lambda^2}{64\pi}\frac{T^2(m_\sigma+T)
(4m_\sigma^2-5m_\pi^2)^2}{6m_\sigma(m_\sigma+T)(4m_\sigma^2-5m_\pi^2)^2+
\pi^2m_\sigma^6(e^{m_\sigma/T}-1)}\,~.
\label{gammahsisi}
\ee

A hard pion in the heat bath can be energetic enough to knock out a sigma 
meson from the condensate. The possible tree-level processes may happen 
according to the four different diagrams. The transition amplitude obtained 
from these is
\be 
{\cal M} &=& -2\lambda\left[1+(m_\sigma^2-m_\pi^2)\left(\frac{1}{s-m_\pi^2} + 
\frac{3}{t-m_\sigma^2} + \frac{1}{u-m_\pi^2}\right)\right]\,~,
\ee
where $s+t+u=2(m_\sigma^2+m_\pi^2)$~. The total scattering cross-section is
\be
\sigma_{\sigma\pi}(s) &=& \frac{\lambda^2}{4\pi s}\left[\left(
\frac{s-2m_\pi^2+m_\sigma^2}{s-m_\pi^2}\right)^2 + \frac{9s(m_\sigma^2-
m_\pi^2)^2}{m_\sigma^2(s^2-sm_\sigma^2-2sm_\pi^2+(m_\sigma^2-m_\pi^2)^2)} 
\right.\nonumber\\
&& \left. +\frac{s(m_\sigma^2-m_\pi^2)^2}{(2m_\sigma^2+m_\pi^2-s)
((m_\sigma^2-m_\pi^2)^2-sm_\pi^2) }\right.\nonumber\\
&&\left. +\frac{6s(m_\sigma^2-m_\pi^2)(sm_\sigma^2+2sm_\pi^2-s^2+m_\sigma^4-
m_\pi^4-2m_\sigma^2m_\pi^2)}{(s-m_\pi^2)(m_\sigma^2+m_\pi^2-s)
(s^2-2s(m_\sigma^2+m_\pi^2)+(m_\sigma^2-m_\pi^2)^2)}\right.\nonumber\\
&& \times\left.\ln{\frac{sm_\sigma^2}{s^2-sm_\sigma^2-2sm_\pi^2+(m_\sigma^2-
m_\pi^2)^2}}\right.\nonumber\\
&& \left. -\frac{2s(m_\sigma^2-m_\pi^2)(3sm_\sigma^2-s^2+m_\pi^4+m_\sigma^4-
4m_\sigma^2m_\pi^2)}{(s-m_\pi^2)(m_\sigma^2+m_\pi^2-s)(s^2-2s(m_\sigma^2+
m_\pi^2)+(m_\sigma^2-m_\pi^2)^2)}\right.\nonumber\\
&& \times\left.\ln{\frac{s(2m_\sigma^2+m_\pi^2-s)}{(m_\sigma^2-m_\pi^2)^2-
sm_\pi^2}}\right] \,~.
\label{crosshsipi}
\ee
The low energy limit is obtained by expanding this about 
$s=(m_\sigma+m_\pi)^2$ and is
\be
\sigma_{\sigma\pi}(s) &=& \frac{9}{4\pi}\lambda^2\frac{m_\pi^4(3m_\sigma^2-
4m_\pi^2)^2}{sm_\sigma^2(m_\sigma^2-4m_\pi^2)^2}\,~.
\label{lowhsipi}
\ee
At high temperatures, where only the four-vertex diagram contributes, the 
cross-section is reduced to 
\be
 \sigma_{\sigma\pi}(s) = \frac{\lambda^2}{4\pi s}\,~.
\ee
The contribution to the imaginary part of the self-energy at low energies is
\be
\mbox{Im}~\Pi_{\sigma\pi} = -\frac{9}{4\pi^3}\lambda^2T^2e^{-m_\pi/T}
F_{\sigma\pi}(m_\sigma,m_\pi)\,~,
\ee
where we defined 
\be
F_{\sigma\pi}(m_\sigma,m_\pi) = \frac{m_\pi^4(m_\pi+T)
(3m_\sigma^2-4m_\pi^2)^2}{m_\sigma^3(m_\sigma+m_\pi)^2
(m_\sigma^2-4m_\pi^2)^2}\,~,
\ee
and at high energies
\be
\mbox{Im}~\Pi_{\sigma\pi} = -\frac{\lambda^2T^2}{96\pi}\,~.
\ee
The two limits can be combined into one approximate expression which then 
determines the rate of dissipation. Factoring in all the $N-1$ pions
\be
\Gamma_{\sigma\pi} \simeq \frac{9(N-1)}{8\pi}\lambda^2\frac{T^2}{m_\sigma}
\frac{F_{\sigma\pi}}{216F_{\sigma\pi}+\pi^2(e^{m_\pi/T}-1)}\,~.
\label{gammahsipi}
\ee
Distinction should be made between scatterings with massless and massive 
pions. The low energy expression (\ref{lowhsipi}) vanishes for zero pion 
mass. The first nonzero term in the series expansion of the cross section is 
the fourth order term. We discuss the massless case in our forthcoming paper 
\cite{future}. 

\begin{figure}[htbp] 
\begin {center}
\vspace*{-3cm}
\leavevmode
\hbox{
\epsfysize=15cm
\epsffile{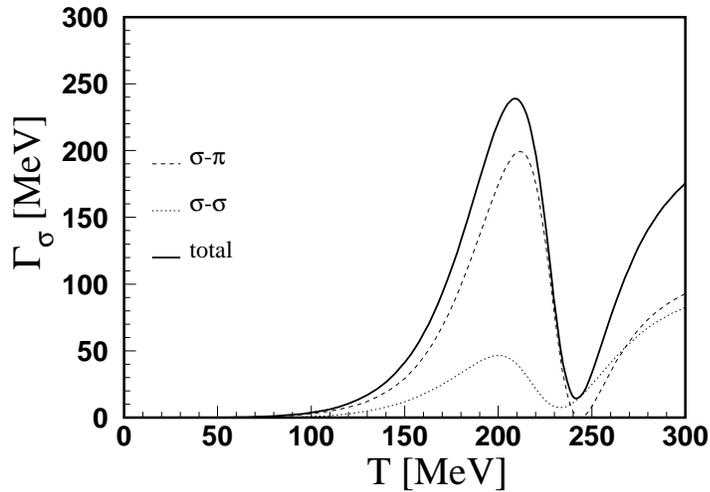}
}
\end{center}
\vspace*{-5cm}
\caption{Scattering contribution to the width of the sigma meson as function 
of temperature. Calculations were done with resummed meson masses. 
\label{hsiscatt.fig}}
\end{figure}

The temperature dependence of the total scattering rate of the sigma meson, 
$\Gamma_{\sigma\sigma}+\Gamma_{\sigma\pi}$, evaluated with the 
self-consistently determined meson masses is shown in figure 
\ref{hsiscatt.fig}. Scattering is more accentuated at higher temperatures and 
its contribution to the sigma damping rates is well below the energy. 
Dissipation of sigmas from the condensate due to their scattering is much 
smaller than due to their decay, meaning that the scalar order parameter 
relaxes to its equilibrium value via the production of lighter pion fields. 
It also means that sigma mesons are so unstable that they are more likely to 
decay before they could ever scatter with other particles from the medium. 
%

\paragraph*{Pion Scattering.}
Dissipation of DCCs can arise from scattering the of soft pions with hard 
pions or hard sigma mesons. The damping of massive pions and that of the 
Goldstone pions we expect to be different, requiring a somewhat different 
analysis. In the following we present the discussion on massive pions. There 
are four possible tree-level diagrams representing the reaction in which a 
hard thermal sigma knocks out a low momentum pion from the condensate. The 
total cross section is the same as for sigma-pion scattering and is given by 
expression (\ref{crosshsipi}). The imaginary part of the self-energy in the 
low energy limit in the rest frame of the pion is 
\be
\mbox{Im}~\Pi_{\pi\sigma} = -\frac{9}{4\pi^3}\lambda^2T^2e^{-m_\sigma/T}
F_{\pi\sigma}\,~,
\ee
where
\be
F_{\pi\sigma} = \frac{m_\pi^5(m_\sigma+T)(3m_\sigma^2-4m_\pi^2)^2}
{m_\sigma^4(m_\sigma+m_\pi)^2(m_\sigma^2-4m_\pi^2)^2}\,~,
\ee
and for high energies is
\be
\mbox{Im}~\Pi_{\pi\sigma} = -\frac{\lambda^2T^2}{96\pi}\,~.
\ee
The total scattering rate due to this process can be parametrized by an 
interpolating formula between the two known limits:
\be
\Gamma_{\pi\sigma} \simeq \frac{9\lambda^2T^2}{8\pi m_\sigma}
\frac{F_{\pi\sigma}}{216F_{\pi\sigma}+\pi^2(e^{m_\sigma/T}-1)}\,~.
\label{gammahpisi}
\ee

$\pi\pi$ scattering has been extensively studied during the last couple of 
decades in a variety of different models and approaches. An incomplete but 
significant list of references is \cite{pipi}. Here we study the elastic 
scattering of a hard pion off a soft pion. For one pion species, $N=2$, there 
are four diagrams contributing to this process at tree-level: one 4-point 
vertex diagram and three 3-point vertex contributions involving a sigma 
exchange in the $s$, $t$ and $u$ channels. The transition amplitude is
\be 
{\cal M} = -2\lambda\left[ 3 + (m_\sigma^2-m_\pi^2)\left(\frac{1}{s-
m_\sigma^2} + \frac{1}{t-m_\sigma^2} + \frac{1}{u-m_\sigma^2}\right)\right]\,~,
\ee
where $s+t+u = 4m_\pi^2$. Accounting for all pion species opens up additional 
channels. For $N=4$ the transition amplitude averaged over initial and summed 
over final isospins is
\be
\left|{\cal M}\right|^2 = \frac{1}{3}\sum_{I=0,1,2}(2I+1)
\left|{\cal M}^I\right|^2\,~,
\ee
where ${\cal M^I}$ is the matrix element associated with the total isospin 
$I$ of the two-pion system:
\be
{\cal M}^0 &=& -2\lambda\left[5 + (m_\sigma^2-m_\pi^2)\left(
\frac{3}{s-m_\sigma^2} + \frac{1}{t-m_\sigma^2} + 
\frac{1}{u-m_\sigma^2}\right)\right]\nonumber\\
{\cal M}^1 &=& -2\lambda (m_\sigma^2-m_\pi^2)\left(
\frac{1}{t-m_\sigma^2} - \frac{1}{u-m_\sigma^2}\right)\nonumber\\
{\cal M}^2 &=& -2\lambda\left[2 + (m_\sigma^2-m_\pi^2)\left(
\frac{1}{t-m_\sigma^2} + \frac{1}{u-m_\sigma^2}\right)\right]\,~.
\ee
The resulting cross-section is
\be
\sigma_{\pi\pi}(s) &=& \frac{\lambda^2}{8\pi s}\left[15 + 10\left(
\frac{m_\sigma^2-m_\pi^2}{s-m_\sigma^2}\right) + 3\left(
\frac{m_\sigma^2-m_\pi^2}{s-m_\sigma^2}\right)^2 +
\frac{6(m_\sigma^2-m_\pi^2)^2}{m_\sigma^2(s+m_\sigma^2-4m_\pi^2)} -\right. 
\nonumber\\
&-&\!\!\!\!\!\!\!\!\!\!\!\!\!\!\!\!\!
\left.\frac{4(m_\sigma^2-m_\pi^2)(5s^2+5sm_\sigma^2-
7m_\sigma^4+13m_\sigma^2m_\pi^2-20sm_\pi^2+4m_\pi^4)}
{(s-4m_\pi^2)(s-m_\sigma^2)(s+2m_\sigma^2-4m_\pi^2)}\ln\left(\frac{s+
m_\sigma^2-4m_\pi^2}{m_\sigma^2}\right)\right]\nonumber\\
\ee
The low-energy limit of the cross-section, quite acceptable for 
$m_\sigma\gg T$, is given by the expansion of this about $s=4m_\pi^2$, 
\be
\sigma_{\pi\pi} = \frac{\lambda^2}{32\pi}\frac{m_\pi^2}{(m_\sigma^2-
4m_\pi^2)^2}\left(23-16\frac{m_\pi^2}{m_\sigma^2}+128\frac{m_\pi^4}
{m_\sigma^4}\right)\,~.
\ee
At high temperatures, in the $T\rightarrow T_c$ limit the major contribution 
to the amplitude is from the four-point vertices, resulting in 
\be
\sigma_{\pi\pi} = \frac{15\lambda^2}{8\pi s}\,~.
\ee
In the rest frame of the pion the above two limits give
\be
\mbox{Im}~\Pi_{\pi\pi} \simeq -\frac{23}{32\pi^3}\lambda^2T^2e^{-m_\pi/T}
\frac{m_\pi^3(m_\pi+T)}{(m_\sigma^2-4m_\pi^2)^2}
\ee
and
\be
\mbox{Im}~\Pi_{\pi\pi} = -\frac{5\lambda^2T^2}{64\pi}\,~,
\ee
respectively. The rate of dissipation due to massive pion-pion scattering is 
then given by the interpolating expression
\be
\Gamma_{\pi\pi} \simeq \frac{23\lambda^2T^2}{64\pi}\frac{m_\pi^2(m_\pi+T)}
{\frac{46}{5}m_\pi^3(m_\pi+T)+\pi^2(m_\sigma^2-4m_\pi^2)^2(e^{m_\pi/T}-1)}\,~.
\label{gammahpipi}
\ee
\begin{figure}[h] 
\begin {center}
\vspace*{-3cm}
\leavevmode
\hbox{
\epsfysize=15cm
\epsffile{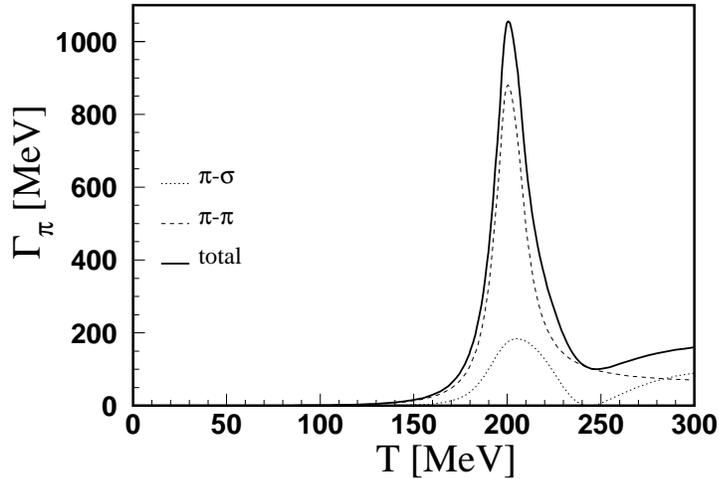}
}
\end{center}
\vspace*{-5cm}
\caption{Scattering contribution to the width of the pion as function of 
temperature. Calculations were done with resummed meson masses.
\label{hpi.fig}}
\end{figure} 

The scattering contribution to the damping of massive pions is presented in 
figure \ref{hpi.fig}. Due to the heavy sigma exchange there is a strong 
suppression at low temperatures. When reaching $T\simeq 130~$MeV the 
scattering rate becomes significant and comparable in magnitude to the 
damping at one-loop order. In the critical region the contribution to the pion
width from pion-pion scattering grows rapidly reaching a maximum about 
$T=200~$MeV. 
It is interesting to note from figures \ref{hsiscatt.fig} and \ref{hpi.fig} 
that in the high temperature regions, where the mesons become almost 
degenerate their widths approach the same value, as expected.

\section{Relaxation Time \label{sect-time}}

\begin{figure}[h] 
\begin {center}
\vspace*{-3cm}
\leavevmode
\hbox{
\epsfysize=15cm
\epsffile{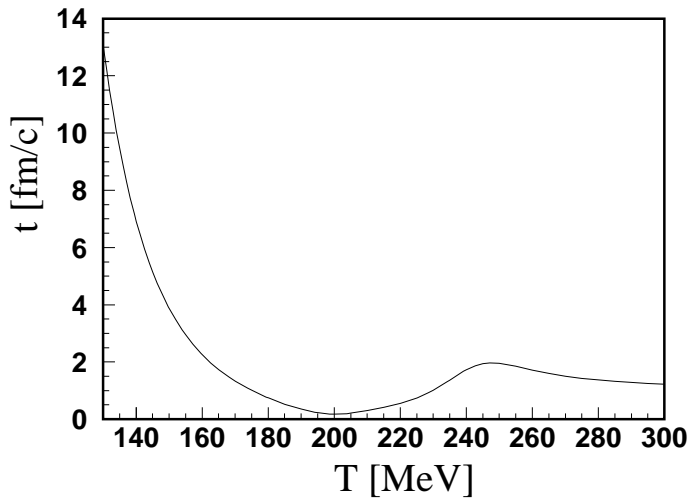}
}
\end{center}
\vspace*{-5cm}
\caption{Relaxation time of a homogeneous disoriented chiral condensate versus 
temperature. \label{time.fig}}
\end{figure}

The ability to detect DCCs in relativistic collisions of heavy ions depends on 
the lifetime of the condensate. DCC formation can happen out of thermal 
equilibrium only. In order to talk about non-equilibrium physics the rate of 
expansion,$t_{exp}$, has to be much smaller than the relaxation time of 
long-wavelength modes, $t$,
\be
t_{exp}\ll t\,~.
\ee
Otherwise, for a slow expansion, the soft modes have enough time to 
equilibrate. Equilibration of the out-of-equilibrium chiral condensate is the 
 result of the presence of a heat bath. Above, we analyzed the physical 
processes responsible, and calculated the damping of different meson modes. 
The total width is the sum of one- and two-loop order dissipative 
contributions,
\be
\Gamma_\pi = \Gamma_{\pi\pi\sigma} + \Gamma_{\pi\sigma} + \Gamma_{\pi\pi}\,~,
\ee
which exhibits a sharp peak in the critical region, due to the peak in the 
damping rate from pion-pion scattering. 
In the non-equilibrium, but close to equilibrium physics that we consider, 
the damping directly controls the rate at which equilibrium is approached 
through the relation \cite{weldon} 
\be
t = \frac{1}{\Gamma_\pi}\,~.
\ee
Obviously, the larger the damping due to the interaction of the condensate 
with the heat bath, the shorter the relaxation time is. Figure \ref{time.fig} 
shows the change in the relaxation time with temperature. 
At low temperatures 
$t$ is large due to the suppression of the thermal occupation numbers. As the 
temperature increases more thermal modes get excited and as a consequence the 
relaxation time is quickly decreasing. In the phase transition region the 
decay time is the shortest. At the peak of the damping $T=200~$MeV we found 
$t=0.17~$fm/c, and at $T=235~$MeV, where the mass of the sigma is the 
smallest, $t=1.36~$fm/c. When assuming no multiple interactions with the heat 
bath then $t$ is the lifetime of the DCC. The time we obtained are shorter 
than previous estimates \cite{rischke,birogreiner,koch}, and are short enough 
to make a possible DCC signal questionable. One can expect that any multiple 
scatterings or decays would only increase the damping, decreasing the 
relaxation time. 

\section{Conclusions \label{sect-conclusions}}

This work was motivated by our interest to determine the possibility of 
survival of DCCs in the background of a multitude of thermal particles, 
mostly pions, that are formed after two heavy ions are collided at 
ultrarelativistic energies. Also, having a consistent description of quantum 
fields near thermal equilibrium allows for a better understanding of their 
dynamics in a region where non-perturbative analysis is required.

We have developed a consistent semi-classical study of the out-of-equilibrium 
chiral condensate fields in the framework of the linear sigma model. Clear 
distinction between the soft non-thermal chiral fields and hard thermal modes 
has been made, accounting also for interactions between these. Motivated by 
the high occupancy of the low momentum modes we 
allowed for their classical treatment. The effect of the other 
degrees of freedom has been taken into account by introducing a heat bath of 
mesons. These thermalised high momentum modes have been accounted for in a 
perturbative manner, improved by the resummation of certain diagrams. We 
derived classical equations of motion for the long-wavelength condensate 
fields coupled to the thermal bath. After integrating out the hard modes 
effective field equations resulted, which completely determine the evolution 
of the chiral condensate in space and time. 

The presence of the slowly varying condensate fields cause deviations in
the equilibrium fluctuations of the thermal fields. We identified these as 
linear response functions, since we are dealing with not too far from 
equilibrium scenarios. We have discussed in details the richness of 
information contained within these response functions: They renormalize the 
equations of motion, modifying the particle properties, and 
give rise to dissipation. 

The temperature dependence of the meson masses and that of the equilibrium 
condensate have been determined numerically in a self-consistent manner, both 
in the chiral limit and for explicitly broken chiral symmetry. In the chiral 
limit a first-order phase transition was found, which is an 
artifact of the model. We show that for a small coupling constant, or large N 
limit, the expected second-order transition is recovered. We have considered 
in some detail the Goldstone boson nature of the pion, proving that when 
properly accounting for the tadpole and sunset diagrams the pion remains 
massless at one-loop level in the symmetry broken phase. Also, the tachyon 
problem of mean-field approximations is eliminated by naturally assuring the 
positivity of the masses at all temperatures. In the more realistic case, in 
which chiral symmetry is explicitly broken by the nonzero quark mass, a 
crossover region was identified. The minimum of the sigma mass is at the 
transition temperature of about $235~$MeV. Above this the masses of the pion 
and the sigma become degenerate and the equilibrium condensate vanishes 
asymptotically, signalling an approximate restoration of chiral symmetry.

Due to possible interactions between different degrees of freedom, those 
of the condensate and those of the heat bath, energy exchange is possible and 
particles can be knocked out or put in the condensate. Direct evaluation of 
the response functions results in the rates for the different processes. We 
have identified these physical processes that are responsible for the 
dissipation of long-wavelength modes of the chiral condensate, and have 
confirmed that at high temperatures not only the damping of the sigmas is 
significant, but also that of the pions. At one-loop level, provided a 
kinematic condition is satisfied, a pion from the condensate can annihilate 
with a pion from the heat bath forming a sigma. The damping due to this 
process becomes stronger with increasing temperature. The width of the pion 
can be as big as $55\%$ of its energy. This result makes us question whether 
one can talk about the pion as a quasiparticle in the phase transition region. 

We emphasize the importance of two-loop calculations. Our results show that 
contribution to the damping rate of the condensate provided by two-particle 
elastic scattering processes can get as important as are decay processes.  
Moreover, while decays happen only when certain kinematic conditions are 
satisfied elastic scatterings have no such restrictions. Therefore two-loop 
processes contribute to the width even in regions where one-loop 
contributions are suppressed. Soft pions from the condensate can be knocked 
out through elastic scatterings with a hard thermal pion or sigma. The damping 
due to these scatterings is most accentuated in the phase transition region, 
and is the greatest at about $200~$MeV, where the pion width shows a peak. 

The damping directly controls the rate at which equilibrium is achieved by the 
non-equilibrium condensate. We have determined the relaxation time of a 
homogeneous condensate due to both one- and two-loop order dissipative 
processes. We have obtained relaxation times between $0.17-1.36~$fm/c in the 
phase transition region. Assuming no multiple interactions these times become 
the lifetime of the condensate. We have found that the lifetime of disoriented 
chiral condensates is short enough to make a possible DCC signal questionable. 

The natural next step of our investigation is to solve the field equations for 
some initial conditions for an expanding system. Such analysis is on the way.  

The methods used in this paper are general, and may be used in other contexts, 
where non-equilibrium physics of quantum fields is of interest. 

\section*{Acknowledgements} 

The author is thankful to J. Kapusta for suggesting working on this problem 
and discussing it at numerous occasions. Thanks to D. Boyanovsky, P.J. Ellis, 
I.N. Mishustin and R. Pisarski for discussions, J.T. Lenaghan and K. Tuominen 
for critically reading and commenting on the manuscript. The financial 
support from The Niels Bohr Institute, the U.S.\ Department of Energy 
Contract No.\ DE-FG-02-91ER-40609, and the L.T.\ Dosdall fellowship of the 
University of Minnesota is acknowledged.  


\end{fmffile}
\end{document}